\documentclass[aps,draft]{revtex4}
\usepackage[dvips]{graphicx}
\begin{document}

\title{Some New Aspects of Degenerate Quantum Plasma}
\author{ Nodar L. Tsintsadze }
\affiliation{Salam Chair in Physics, GC University, Lahore-54000, Pakistan\\Department of Plasma Physics, E.Andronikashvii Institute of Physics, Tbilisi, Georgia }

\date{\today}

\begin{abstract}
 Answers to some salient questions, which  arise in quantum plasmas, are given. Starting from  the Schr\"{o}dinger equation for  a single  particle it is demonstrated how the Wigner-Moyal equation can be derived. It is shown that the Wigner-Moyal type of equation  also exists in  the classical field theory. As an example, from  the Maxwell equations the Wigner-Moyal type of equation is obtained for a dense photon gas, which is classical, concluding that the Wigner-Moyal type of equation  can be derived for any system, classical or quantum. A new type of quantum kinetic equations are presented. These novel kinetic equations allow to obtain a set of quantum hydrodynamic equations, which is impossible to derive by the Wigner-Moyal equation. The propagation of small perturbations and instabilities of these perturbations are then discussed, presenting new modes of quantum plasma waves. In the case of low frequency oscillations with ions, a new Bogolyubov type of spectrum is found. Furthermore, the Korteweg-de Vries (KdV) equation is derived and the contribution of the Madelung term in the formation of the KdV solitons is discussed.
\end{abstract}


\maketitle

\section{Introduction}

Our understanding of the collective behavior of quantum plasmas has recently undergone some appreciable theoretical progress \cite{tsin},\cite{tsin10}, \cite{pad}. Quantum plasma is of primary interest in its potential application in modern technology, such as metallic and semiconductor nanostructures. As well as, the quantum plasmas are prevalent  in planetary interiors and in compact astrophysical objects, e.g., the interior of white dwarf stars, magnetospheres of neutron stars and magnetars, etc. Despite extensive theoretical efforts in the remote past \cite{gol}-\cite{bp}, where the properties of linear electron oscillations in a dense Fermi plasma have been studied, and more recent ones \cite{man},\cite{mar}, there are questions and issues which has to be clarified.

The purpose of the present work is to elucidate and give the correct formulation to quantum plasmas, and demonstrate which equations can be used, as well as which are invalid for the description of the collective behavior of quantum plasmas. So that in this paper, we provide answers to some salient questions, which  arise in quantum plasmas. We start from  the Schr\"{o}dinger equation for  a single  particle and demonstrate how the Wigner-Moyal equation can be derived. We then show that the Wigner-Moyal type of equation  also exists in  the classical field  theory. Next we present a new type of quantum kinetic equations. These novel kinetic equations allow us to obtain a set of quantum hydrodynamic equations, which is impossible to derive by the Wigner-Moyal equation. We then discuss the propagation of small perturbations and instabilities of these perturbations, presenting new modes of quantum plasma waves. Finally, we derive the Korteweg-de Vries (KdV) equation in order to discuss the solitary waves in quantum plasmas.

\section{Derivation of the Wigner-Moyal Equation }

We start with a single particle and for this purpose, we employ the non-relativistic Schr\"{o}dinger equation
\begin{eqnarray}
\label{sh}
i\hbar\frac{\partial\Psi(\vec{r},t)}{\partial t}+\frac{\hbar^2}{
2m}\Delta \Psi(\vec{r},t)-V(\vec{r},t)\Psi(\vec{r},t)=0 \ ,
\end{eqnarray}
where $\Psi(\vec{r},t)$ is not only the wave function of the single particle, but also an amplitude of the probability. Namely, $\mid\Psi(\vec{r},t)\mid^2d\vec{r}$ is the probability of finding   values of coordinates of the particle within the interval $\vec{r}, \vec{r}+d\vec{r}$. The probability density is $\mid\Psi(\vec{r},t)\mid^2$, and $\mid\Psi(\vec{r},t)\mid^2\sim\frac{1}{V}$.

Note that the Schr\"{o}dinger equation (\ref{sh}) describes the wave properties of the single particle. For the dense system, N-body ($N\rightarrow\infty$) system with an interaction potential $V(\vec{r}^N)=V(\vec{r}_1,\vec{r}_2,...\vec{r}_N)$ the Schr\"{o}dinger equation reads
\begin{eqnarray}
\label{shnb}
i\hbar\frac{\partial\Psi(\vec{r}^N,t)}{\partial t}=\left[ -\frac{\hbar^2}{2m}\sum_{k=1}^N\ \frac{\partial^2}{
\partial \vec{r}_k^2}+V(\vec{r}^N)\right] \Psi(\vec{r}^N,t)\ .
\end{eqnarray}
Obviously, in this case it is very difficult to discuss the wave properties. Therefore, we should proceed from waves to particles.

We consider Eq.(\ref{sh}) at two different points, $\vec{r}_1$ and $\vec{r}_2$
\begin{eqnarray}
\label{fr1}
i\hbar\frac{\partial\Psi(\vec{r}_1,t)}{\partial t}+\frac{\hbar^2}{
2m}\Delta_1 \Psi(\vec{r}_1,t)-V(\vec{r}_1,t)\Psi(\vec{r}_1,t)=0 \ ,
\end{eqnarray}
\begin{eqnarray}
\label{fr2}
-i\hbar\frac{\partial\Psi^*(\vec{r}_2,t)}{\partial t}+\frac{\hbar^2}{
2m}\Delta_2 \Psi^*(\vec{r}_2,t)-V(\vec{r}_2,t)\Psi^*(\vec{r}_2,t)=0 \ .
\end{eqnarray}
We now multiply Eq.(\ref{fr1}) by $\Psi^*(\vec{r}_2,t)$ and Eq.(\ref{fr2}) by $\Psi(\vec{r}_1,t)$, and subtract the resulting equations to obtain
\begin{eqnarray}
\label{sto}
i\hbar\frac{\partial}{\partial t}\ \Psi(\vec{r}_1,t)\Psi^*(\vec{r}_2,t)+\frac{\hbar^2}{
2m}(\Delta_1-\Delta_2) \Psi(\vec{r}_1,t)\Psi^*(\vec{r}_2,t)-[V(\vec{r}_1,t)-V(\vec{r}_2,t)]\Psi(\vec{r}_1,t)\Psi^*(\vec{r}_2,t)=0 \ .
\end{eqnarray}

Introducing a new variables $\vec{\rho}=\vec{r}_1-\vec{r}_2,\ $ $\ \vec{r}=\frac{1}{2}(\vec{r}_1+\vec{r}_2)$, so that
\begin{eqnarray*}
\Delta_1-\Delta_2=\Bigl(\frac{\partial}{\partial \vec{r}_1}-\frac{\partial}{\partial \vec{r}_2}\Bigr)\Bigl(\frac{\partial}{\partial \vec{r}_1}-\frac{\partial}{\partial \vec{r}_2}\Bigr)=2\frac{\partial^2}{\partial \vec{r}\partial\vec{\rho}}\ ,
\end{eqnarray*}
\begin{eqnarray*}
\Psi(\vec{r}_1,t)\Psi^*(\vec{r}_2,t)=\Psi(\vec{r}+\frac{\vec{\rho}}{2},t)\Psi^*(\vec{r}-\frac{\vec{\rho}}{2},t)\ ,
\end{eqnarray*}
and multiplying Eq.(\ref{sto}) by $e^{-\frac{i\vec{p}\vec{\rho}}{\hbar}}$, and integrating over $\vec{\rho}$ we get
\begin{eqnarray}
\label{ex}
\frac{\partial f}{\partial t}+\frac{(\vec{p}\cdot\nabla)}{m}f+\frac{1}{i\hbar}\int d\vec{\rho}\Bigl\{V\Bigl(\vec{r}+
\frac{\vec{\rho}}{2},t\Bigr)-V\Bigl(\vec{r}-\frac{\vec{\rho}}{2},t\Bigr)\Bigr\}e^{-\frac{i\vec{p}\vec{\rho}}{\hbar}}\ \Psi(\vec{r}+\frac{\vec{\rho}}{2},t)\Psi^*(\vec{r}-\frac{\vec{\rho}}{2},t)=0\ ,
\end{eqnarray}
where
\begin{eqnarray}
\label{wf}
f=f_w=\int d\vec{\rho}e^{-\frac{i\vec{p}\vec{\rho}}{\hbar}}\ \Psi(\vec{r}+\frac{\vec{\rho}}{2},t)\Psi^*(\vec{r}-\frac{\vec{\rho}}{2},t)
\end{eqnarray}
is the Wigner function.

In order to derive the Wigner-Moyal equation  one should suppose that  $V(\vec{r},t)$ is a smooth function.  Expansion $V\Bigl(\vec{r}+\frac{\vec{\rho}}{2},t\Bigr)-V\Bigl(\vec{r}-\frac{\vec{\rho}}{2},t\Bigr)$ in Taylor series leads to the Wigner-Moyal equation
\begin{eqnarray}
\label{wme}
\frac{\partial f_w}{\partial t}+\frac{(\vec{p}\cdot\nabla)}{m}f_w-\frac{2}{\hbar}sin\Bigl(\frac{\hbar}{2}\nabla_{\vec{r}}\cdot\nabla_{\vec{\rho}}
\Bigr)\ V(\vec{r},t)f_w=0\ .
\end{eqnarray}
It should be emphasized that this equation has the following restrictions: $V(\vec{r},t)$ should be the smooth function, and Eq.(\ref{wme}) is not suitable for the construction of fluid equations, even if we apply it to a macroscopical body.

Note that the integration of the Wigner function over momenta gives the configurational  probability density
\begin{eqnarray}
\label{cpd}
\int f_w d\vec{p}=\mid\Psi(\vec{r},t)\mid^2\ ,
\end{eqnarray}
whereas the integration over spatial coordinates gives the momentum probability
\begin{eqnarray}
\label{mprob}
\int f_w d\vec{r}=\mid\varphi(\vec{\rho},t)\mid^2\ .
\end{eqnarray}
Moreover, the mean value of any physical quantity $L(p,q)$ can be calculated by the statistical distribution function
\begin{eqnarray}
\label{sdf}
<L>=\int L(p,q)f_w(\vec{r},\vec{p})d\vec{r}d\vec{p}\ .
\end{eqnarray}
However, the Feynman notes \cite{fey} "Although Wigner function $f_w(\vec{r},\vec{p})$ satisfies Eqs.(\ref{cpd}) and (\ref{mprob}), it cannot be regarded as the probability for finding the particle at the point $\vec{r}$ and with the momentum $\vec{p}$, because $f_w(\vec{r},\vec{p})$ can become negative for some values of $\vec{p}$ and $\vec{r}$. Also, Eq.(\ref{sdf}) is not true for a general function $L(p,q)$". Thus the Wigner distribution function has no physical meaning.

We here also write the Wigner distribution function for N-particle system
\begin{eqnarray*}
f_w(\vec{r}^N,\vec{p}^N,t)=\int d\vec{\rho}\ e^{-\frac{i\vec{p}^N\vec{\rho}^N}{\hbar}}\ \Psi(\vec{r}^N+\frac{\vec{\rho}^N}{2},t)\Psi^*(\vec{r}^N-\frac{\vec{\rho}^N}{2},t)\ .
\end{eqnarray*}

The question is: does the Wigner's type of equation exist in the classical approximation, or it is purely quantum?
We emphasize that such type of equation one can derive for any system, classical or quantum. To show this, we can employ  the relativistic Maxwell's equation
\begin{eqnarray}
\label{rme}
\Delta\vec{A}-\frac{\partial^2\vec{A}}{\partial t^2}=\frac{n}{\gamma}\vec{A} \hspace{1.5cm} where
\hspace{1.5cm} \gamma=\sqrt{1+\frac{e^2A^2}{m^2c^4}}
\end{eqnarray}
and consider it at two distinct points $(\vec{r}_1,\vec{r}_2)$. Following the procedure described above for the Wigner-Moyal  equation, one can obtain for the occupation number $N(\vec{r},\vec{k},t)$ of photons \cite{ltsin98},\cite{ntsin98},\cite{men}
\begin{eqnarray}
\label{ocnu}
\frac{\partial}{\partial t}N(\vec{r},\vec{k},t)+(\vec{v}_g\cdot\nabla)N(\vec{r},\vec{k},t)-\omega_p^2sin\Bigl(\frac{1}{2}\nabla_{\vec{r}}\cdot
\nabla_{\vec{k}}\Bigr)V(\vec{r},t)N(\vec{r},\vec{k},t)=0
\end{eqnarray}
and
\begin{eqnarray*}
N_\gamma=<N>=\int\ \frac{d\vec{k}}{(2\pi)^3}N(\vec{r},\vec{k},t)>>1\ ,
\end{eqnarray*}
which describes the dense photon gas.

\section{New Kinetic Equations }

Some uncertainty of the Wigner-Moyal equation demands the derivation of a new quantum kinetic equation incorporating  all the quantum properties of the body. Moreover, from which must follow a set of fluid equations with a quantum term. To this end, we start with a single fermi particle and  employ the
non-relativistic Pauli equation \cite{ber}, which reads
\begin{eqnarray}
\label{pau}
i\hbar \frac{\partial \Psi_\alpha}{\partial t}+\frac{\hbar^2}{
2m_\alpha}\Delta \Psi_\alpha -\left[ \frac{ie\hbar }{2m_\alpha c}
(\vec{A}\cdot\nabla +\nabla\vec{A})+\frac{e^2A^2}{2m_\alpha c^2}
+e_\alpha \varphi -\vec{\mu }_\alpha \cdot\vec{H}\right] \Psi _\alpha =0 \ ,
\end{eqnarray}
where $\Psi_{\alpha }=\Psi _{\alpha }(\vec{r},t,\vec{\sigma} )$ is the wave function of the single particle species $\alpha $, having the spin $\vec{s}=1/2\vec{\sigma }$ ($\sigma=\pm 1$). $\vec{A}(\vec{r},t)$
and $\varphi (\vec{r},t)$ are the vector and scalar potentials, respectively. The last
term in Eq.(\ref{pau}) is the potential energy of magnetic dipole in the external
magnetic field, the magnetic moment of which is
\begin{eqnarray}
\label{mm}
\vec{\mu}_\alpha =\frac{e\hbar }{2m_\alpha c}
\vec{\sigma }=\mu_\beta \vec{\sigma } \ ,
\end{eqnarray}
where $\mu_{\beta }$ is the Bohr magneton and $\vec{\sigma}$ is the operator of the single particle \cite{ber},\cite{lan}.

Use of the Madelung representation \cite{mad} of the complex function $\Psi_\alpha$
\begin{eqnarray}
\label{mad}
\Psi_\alpha (\vec{r},t,\vec{\sigma} )=a_\alpha (\vec{r},t,\vec{\sigma} )\exp \frac{iS_\alpha (\vec{r},t,\vec{\sigma} )}{\hbar }\ ,
\end{eqnarray}
where $a_\alpha (\vec{r},t,\vec{\sigma} )$ and $S_\alpha (\vec{r},t,\vec{\sigma} )$ are real, in
the Pauli equation (\ref{pau}), yields the following two equations
\begin{eqnarray}
\label{for}
\frac{\partial a_\alpha^2(\vec{r},t,\vec{\sigma} )}{\partial t}+\nabla \cdot
\Bigl(a_\alpha^2(\vec{r},t,\vec{\sigma} )\frac{\vec{p}_\alpha }{m_\alpha }\Bigr)=0 \ ,
\end{eqnarray}
\begin{eqnarray}
\label{mom}
\frac{d\vec{p}_\alpha }{dt}=e_\alpha \Bigl(\vec{E}+\frac{\vec{v}_\alpha \times \vec{H}}{c}\Bigr)+\frac{
\hbar^2}{2m_\alpha }\nabla \frac{1}{a_\alpha }\nabla^2a_\alpha +
\mu_\beta\nabla (\vec{\sigma}\cdot\vec{H})\ .
\end{eqnarray}
Equation (\ref{for}) has an obvious physical meaning. Namely, $a_\alpha^2=|\Psi_\alpha|^2$ is the probability density of finding the single particle at some point in space with a spin $\vec{s}$. Whereas, $\vec{p}_\alpha=\nabla \vec{S}-\frac{e_\alpha}{c}\vec{A}$ is the momentum operator of the particle.
Note that if there is no spin dependence of the wave functions $\Psi_\alpha(\vec{r},t)$, then $\vec{p}_\alpha(\vec{r},t)$ becomes the ordinary momentum of the particle.
It should be emphasized that in Eq.(\ref{mom})  the second, quantum Madelung, term describes the diffraction
pattern of a single electron \cite{bib}.

In the case when there are no external electric and magnetic
fields ($\vec{E}=0$, $\vec{H}=0$), after the linearization
of Eqs. (\ref{for}) and (\ref{mom}), we get the frequency of quantum oscillations of
a free electron
\begin{eqnarray}
\label{fqo}
\omega_q=\frac{\hbar k^2}{2m}\ .
\end{eqnarray}

Based on the diffraction pattern of electrons, Born has given a statistical interpretation of the wave function, which states that
in every point in space at a given time the intensity of the de Broglie waves is proportional to the probability of observing a particle at that point in space. Therefore,  a density of probability distribution $f_S$ in phase space for the single particle is
\begin{eqnarray}
\label{prob}
|\Psi |^2=\int d^3p\ f_S(\vec{r},\vec{p},t)\ .
\end{eqnarray}
This function $f_S$, obviously must satisfy the normalization condition over all phase space
\begin{eqnarray}
\label{nor}
\int d^3r\int d^3p\ f_S=\int d^3r\ |\Psi |^2=1\ .
\end{eqnarray}
Moreover, Bogolyubov has introduced a one particle distribution function \cite{bog} for the system as a whole from Liouville's theorem regarding the distribution function $f_\alpha^N(t,\tau_1,\tau_2...\tau_N),$ (where $\tau_\alpha$ is the set of coordinates and momentum components for the $\alpha$ particle), and derived the Vlasov and Boltzmann equations in the gas approximation, which means that the plasma parameter $\eta $ (representing the ratio of the average potential energy $<U>$ of particles interaction to the average kinetic energy $<\varepsilon_k>$) must be less than unity, i.e., $\eta =\frac{<U>}{<\varepsilon_k>}\ll 1$.
Note that the one particle distribution function $f(\vec{r},\vec{p},t)$ is normalized to unity, whereas the
Liouville's function $f^N$ to total number of particles, i.e., $f^N=Nf(\vec{r},\vec{p},t)$. The same relation between $f^N$ and $f(\vec{r},\vec{p},t)$ in an alternative description of kinetic theory was obtained by Klimontovitch \cite{kli},\cite{lib}. To make it more lucid, we shall give a simple explanation about a single particle and one particle distribution function \cite{lib},\cite{lan}. Namely, the probability density of the single particle is one particle per unit volume, $n^s(\vec{r},t)=|\Psi (\vec{r},t)|^2$, with dimensions 1/V. Whereas the one particle distribution function means that in spite of the large number of particles
in the unit volume all of them have only one $\vec{r}$ and $\vec{p}$. This permits us to express the number density of particles per unit volume as $n(\vec{r},t)=\int d^3p\ f(\vec{r},\vec{p},t)=N/V$. Therefore, we can write
\begin{eqnarray}
\label{ns}
n(\vec{r},t)=Nn^s(\vec{r},t)=N|\Psi (\vec{r},t)|^2=\int d^3p\ f(\vec{r},\vec{p},t)\ .
\end{eqnarray}
Thus, $n(\vec{r},t)$ is the density of quantum particles per unit volume. We
have assumed that the total number of particles of each species is conserved.

Non-equilibrium states of a Fermi quantum gas are described by the one particle
distribution function $f_\alpha (\vec{r},\vec{p},t,\vec{\sigma} ),$ which satisfies the
quantum Boltzmann equation
\begin{eqnarray}
\label{qb}
\frac{\partial f_\alpha (\vec{r},\vec{p},t,\vec{\sigma} )}{\partial t}+\left( \frac{\partial
\vec{r}}{\partial t}\cdot\nabla\right) f_\alpha (\vec{r},\vec{p},t,\vec{\sigma} )+\frac{d\vec{p}_\alpha }{dt}\frac{\partial
f_\alpha (\vec{r},\vec{p},t,\vec{\sigma} )}{\partial \vec{p}}=C(f_\alpha ) \ .
\end{eqnarray}
Equation (\ref{qb}) for quasi-particles in a Fermi liquid was written by Landau \cite{lanj},\cite{lif}.
Here $C(f_\alpha )$ is the collision integral, which describes the
variation of the distribution function due to particle collisions, the
derivative $\frac{dp_\alpha }{dt}$ is determined by the force acting on
the particle, the expression of which is given by the equation (\ref{mom}). When
the spin of particles is taken into account, the distribution function $f_\alpha $ becomes an operator with respect to the spin variables $\sigma $. In this case the total number density of particles $n_\alpha (\vec{r},t)$ equals
\begin{eqnarray}
\label{tnd}
n_\alpha (\vec{r},t)=\sum_\sigma \int \frac{d^3p}{(2\pi \hbar )^3}f_\alpha (\vec{r},\vec{p},t,\vec{\sigma})\ .
\end{eqnarray}
If there is no spin dependence of the distribution function, then $f_\alpha $ becomes the ordinary quasi-classical distribution function $f_\alpha (\vec{r},\vec{p},t)$. Note that the condition for quasi-classical motion is that the de Broglie wavelength $\lambda_B=\hbar/p_F$ of the particle must be very small compared with the characteristic length L, over
which $f(\vec{r},\vec{p},t)$ varies considerably.

If collisions between particles were entirely negligible, each particle of
the system would constitute a closed subsystem, i.e., one could neglect the
collision integral in Eq.(\ref{qb}), and the distribution function of particles would
obey Liouville-Vlasov equation.

Hereafter, we consider the system as a spinless.
We substitute the equation of motion of single particle (\ref{mom})
(neglecting the last term) into the kinetic equation (\ref{qb}) taking into account the
definition of the density of particles (\ref{ns}) to obtain
\begin{eqnarray}
\label{tob}
\frac{\partial f_\alpha }{\partial t}+\left( \vec{v}\cdot\nabla \right) f_\alpha
+e_\alpha \Bigl(\vec{E}+\frac{\vec{v}_\alpha \times \vec{H}}{c}\Bigr)\frac{\partial f_\alpha }{
\partial \vec{p}}+\frac{\hbar^2}{2m_\alpha }\nabla \frac{1}{\sqrt{
n_\alpha }}\Delta \sqrt{n_\alpha }\ \frac{\partial f_\alpha }{\partial \vec{p}}=C(f_\alpha )\ .
\end{eqnarray}
It should be emphasized that this is a novel equation with the quantum term, which contains all the information on
the quantum effects. We specifically note also that this equation is rather simple from the mathematically point of
view.

\section{Proper Waves in a Fermi Quantum Plasma}

We now employ the equation (\ref{tob}) to study the propagation of small longitudinal
perturbations ($\vec{H}=0$, $\vec{E}=-\nabla \varphi $) in an electron-ion collisionless plasmas.
For a weak field, we look for the electron and ion distribution functions in the form $f_\alpha =f_{\alpha 0}+\delta f_\alpha $, where $f_{\alpha 0}$ is the stationary isotropic homogeneous distribution function unperturbed by the field, and $\delta f_\alpha $ is the small variation in it due to field.
After linearization of Eq.(\ref{tob}) with respect to the perturbation, we assume $\delta f_\alpha $ and $\delta \varphi $ vary like $expi(\vec{k}\cdot\vec{r}-\omega t)$.

Using the Poisson's equation
\begin{eqnarray}
\label{puas}
\Delta \delta \varphi =4\pi e\left\{ 2\int \frac{d^3p}{(2\pi \hbar )^3}
\delta f_e-2\int \frac{d^3p}{(2\pi \hbar )^3}\delta f_i\right\}
\end{eqnarray}
and assuming the Fermi degeneracy temperature $T_{F}=\frac{\varepsilon_F}{K_B}$ ($K_B$ is the Boltzmann coefficient, the Fermi distribution function is the step function $f_{\alpha 0}=\Theta (\varepsilon_{F\alpha }-\varepsilon )$, where
$\varepsilon_{F\alpha }=\frac{m_\alpha v_{F\alpha }^2}{2}\ $) much more than the Fermi gas temperature,
then we obtain after some algebra the quantum dispersion equation
\begin{eqnarray}
\label{qde}
\varepsilon =1+\sum_\alpha\frac{3\omega_{p\alpha}^2}{\Gamma_\alpha k^2v_{F\alpha}^2}\left\{1-\frac{\omega }{2kv_{F\alpha}}\ln
\frac{\omega +kv_{F\alpha}}{\omega -kv_{F\alpha}}\right\}=0 \ ,
\end{eqnarray}
where
\[
\Gamma_\alpha=1+\frac{3\hbar^2k^2}{4m_\alpha^2 v_{F\alpha}^2}\Bigl(1-\frac{\omega }{
2kv_{F\alpha}}\ln \frac{\omega +kv_{F\alpha}}{\omega -kv_{F\alpha}}\Bigr)\ ,
\]
and $\omega$ can be more or less than $kv_{Fe}$. Note that for $\omega \gg kv_{Fi},\ $ $\Gamma_i\approx 1.$

Let us first consider the electron Langmuir waves, supposing that the ion
mass $m_i\rightarrow \infty $ and $\omega \gg kv_{Fe},\ $ or the
range of fast waves, when the phase velocity exceeds the Fermi velocity of
electrons. In this case we get the dispersion relation
\begin{eqnarray}
\label{dlw}
\omega^2=\omega_{pe}^2+\frac{3k^2v_{Fe}^2}{5}+\frac{\hbar^2k^4}{4m_e^2}+...
\end{eqnarray}
which has been previously derived by Klimontovich and Silin \cite{klis}. The expression
(\ref{dlw}) exhibits that the high frequency oscillations of electrons of a degenerate
plasma remain undamped in the absence of particles collisions. Note that the
Landau damping is also absent, since according to the Fermi distribution there
are no particles with velocities greater than the Fermi velocity which could
contribute to the absorption.

As k increases, Eq.(\ref{dlw}) becomes invalid, but still $\omega > kv_{Fe}$ and the Landau damping is absent. We now introduce the Thomas-Fermi screening wave vector $k_{TF}=\frac{\sqrt{3}\omega_{pe}}{v_{Fe}}.\ $ In the limit $k^2\gg k_{TF}^2,\ $ $\omega $ tends to $kv_{Fe}\ $ (at $\ m_i\rightarrow \infty )\ $ and we obtain from (\ref{qde})
\begin{eqnarray}
\label{klim}
\omega =kv_{Fe}\Bigl(1+2\exp \{-\frac{2(\frac{k^2}{k_{TF}^2}+
\frac{\omega_q^2}{\omega_{pe}^2})}{1+\frac{\omega_q^2}{\omega_{pe}^2}}\}\Bigr)\ .
\end{eqnarray}
If we neglect the quantum term $\omega_q$ in Eq.(\ref{klim}), then
we recover waves known as the zero sound, which are the continuation of the
electron Langmuir wave (\ref{dlw}) into the range of short wavelength. Thus the
expression (\ref{klim}) represents the quantum correction to the zero sound.

Special and very important case in a quantum plasma is an one-fluid
approximation. In this case one assumes that the characteristic dimension R of inhomogeneities  in the plasma is larger than the electron Thomas-Fermi length $\lambda_{TF}=\frac{v_{Fe}}{\sqrt{3}\omega_{pe}}\ $. In the perturbed plasma the potential of electric field is determined by  the Poisson`s equation
\begin{eqnarray}
\label{tpo}
\Delta \varphi =4\pi e(\delta n_e-Z_i\delta n_i)\ .
\end{eqnarray}
To estimate a magnitude of the term on the right-hand side of Eq.(\ref{tpo}), we consider two cases $e\varphi\ll \varepsilon_F$ and $e\varphi\sim \varepsilon_F$.
Noting that $\Delta \varphi\sim\varphi/R^2$, for the weak electric field $\delta n_e=\frac{3}{2}n_0\frac{e\varphi}{\varepsilon_F}$
we rewrite Eq.(\ref{tpo}) in the form
\begin{eqnarray}
\label{npo}
\mid\frac{\delta n_e-Z_i\delta n_i}{\delta n_e}\mid=\frac{\lambda_{TF}^2}{R^2}\ll 1 \ .
\end{eqnarray}
This inequality (\ref{npo}) is also satisfied for the strong perturbation, $e\varphi\sim \varepsilon_F$ and $\delta n_e\sim n_e$.
Thus we can conclude that the uncompensated charge density is small compared to the perturbation of the electron and ion charge density separately. So, for $\lambda_{TF}\ll R$ we further assume that the quasi neutrality
\begin{eqnarray}
\label{qneu}
n_e=n_i
\end{eqnarray}
is satisfied. This equation along with the equation of motion of ions and the
equation giving the adiabatic distribution of electrons allow us to define
the potential field. In such approximation the charge is completely
eliminated from the equations, and the Thomas-Fermi length $\lambda_{TF}$ disappears with it.

In order to construct the one-fluid quantum kinetic equation, we neglect the
time derivative in the equation (\ref{tob}) of electrons, as well as the collision terms, suppose
$\vec{E}=-\nabla \varphi$ and $\vec{H}=0$,
and write the dynamic equations for the quasi-neutral plasma (\ref{qneu})
\begin{eqnarray}
\label{deq1}
(v\cdot\nabla )f_e+\nabla \Bigl(e\varphi +\frac{\hbar^2}{2m_e}\frac{1}{\sqrt{n}}
\Delta \sqrt{n}\Bigr)\frac{\partial f_e}{\partial p}=0\ ,
\end{eqnarray}
\begin{eqnarray}
\label{deq2}
\frac{\partial f_i}{\partial t}+(v\cdot\nabla )f_i-\nabla e\varphi \ \frac{
\partial f_i}{\partial p}=0\ .
\end{eqnarray}
In Eq.(\ref{deq2}) we have neglected the quantum term as a small one. Note that the Fermi distribution function of electrons
\begin{eqnarray}
\label{fde}
f_e=\frac{1}{\exp \left\{\frac{\frac{p^2}{2m_e}-U-\mu_e}{T}\right\}+1}
\end{eqnarray}
satisfies the equation (\ref{deq1}). Here $U=e\varphi +\frac{\hbar^2}{2m_e}\frac{1}{\sqrt{n}}\Delta \sqrt{n}\ ,$
and $\mu_e$ is the chemical potential.

For the strongly degenerate electrons, i.e., $T_e\rightarrow 0$ ($\mu_e=\varepsilon_F$), the Fermi distribution function
becomes the step function
\begin{eqnarray}
\label{stf}
f_e=\Theta (\varepsilon_F+U-\frac{p^2}{2m_e})\ ,
\end{eqnarray}
which allows us to define the density of electrons ($n_e=n_i=n=2\int\frac{d^3p}{(2\pi\hbar )^3}f $)
\begin{eqnarray}
\label{del}
n=\frac{p_{Fe}^3}{3\pi^2\hbar^3}\left( 1+\frac{e\varphi +\frac{
\hbar^2}{2m_e }\frac{1}{\sqrt{n }}\Delta \sqrt{n}}{\varepsilon _{Fe}}\right)^{3/2}\ .
\end{eqnarray}

We now express $e\varphi $ from the equation (\ref{del}) and substitute it into the
kinetic equation (\ref{deq2}) to obtain
\begin{eqnarray}
\label{ofk}
\frac{\partial f}{\partial t}+(v\cdot\nabla )f-\nabla \left\{ \varepsilon
_{Fe}\left(\frac{n}{n_0}\right)^{2/3}-\frac{\hbar^2}{2m_e}\frac{1}{\sqrt{n}}
\Delta \sqrt{n}\right\} \frac{\partial f}{\partial p}=0\ .
\end{eqnarray}
This is the nonlinear kinetic equation of quantum plasma in the one-fluid approximation, which incorporates the potential energy due to degeneracy of the plasma and the Madelung potential.

To consider the propagation of small perturbations $f=f_0(\vec{p})+\delta f(\vec{r},\vec{p},t)$ and $
n=n_0+\delta n(\vec{r},t),$ we shall linearize Eq.(\ref{ofk}) with respect to the perturbations, look for a
plane wave solution as $e^{i(\vec{k}\cdot\vec{r}-\omega t)},$ and derive the
dispersion equation, which resembles the Bogolyubov's dispersion
relation in the frequency range $kv_{Fe}\gg\omega\gg kv_{Fi},$
\begin{eqnarray}
\label{bog}
\omega =k\sqrt{\frac{p_F^2}{3m_e m_i}+\frac{\hbar^2k^2}{
4m_i m_e}}=k\sqrt{\frac{2\varepsilon_{Fe}}{3m_i}+\frac{\hbar^2k^2}{4m_i m_e}}\ .
\end{eqnarray}
We specifically note here that this type of spectrum (\ref{bog}) was derived by Bogolyubov for the elementary
excitations in a quantum Bose liquid, and the microscopic theory
of the superfluidity of liquid helium was created \cite{lif},\cite{bogj}.

How one can explain the similar spectrum (\ref{bog}) in the Fermi gas? In the dense
and low temperature plasma the formation of bound states is possible due to the attractive character of the Coulomb force
\cite{lan},\cite{lans},\cite{pin},\cite{kre}. As is well known, when the density of particles increases and the temperature goes to zero,
the nuclear reaction leads to the capture of electrons by nuclei. In such
reaction the charge on the ions (nucleus) decreases. Because we have assumed
the quasi-neutrality, we have therefore supposed that all electrons are in the bound state with ions.
Thus one can say that in the one-fluid approximation the Fermi
plasma may become the Bose system due to the bound state or this approximation may imply the formation of the Bose atoms.

The same Bogolyubov's type of spectrum (\ref{bog}) follows from Eq.(\ref{qde}) in the range of
intermediate phase velocities $kv_{Fe}\gg\omega\gg kv_{Fi}.$ In this case, except the expression for the real part of frequency (\ref{bog}), we get the imaginary part of $\omega $ from the dispersion relation (\ref{qde})
\begin{eqnarray}
\label{im}
Im\omega =-\frac{\pi }{12}k\ \frac{p_{Fe}}{m_i} \ ,
\end{eqnarray}
which is much less than $Re\omega .$ The expression (\ref{im}) indicates that the electrons play role in
the absorption of oscillations, since their random
velocities greatly exceed the phase velocity. So that the damping rate
is determined by the electrons alone.

\section{Derivation of Quantum Hydrodynamic Equations}

We shall now derive a set of fluid equations. The Boltzmann and Vlasov type of quantum kinetic equations (\ref{tob}) and (\ref{ofk}) give a microscopic description of the way in which the state of the plasma varies
with time. It is also well known how the kinetic equation can be
converted into the usual equations of fluids. Following the standard method, we can derive the equations of continuity  and motion of macroscopic quantities from Eq.(\ref{ofk})
\begin{eqnarray}
\label{con}
\frac{\partial n}{\partial t}+\nabla\cdot(n\vec{u})=0
\end{eqnarray}
\begin{eqnarray}
\label{mot}
\frac{\partial \vec{u}}{\partial t}+(\vec{u}\cdot \nabla )\vec{u}=-\frac{
K_{B}T_{Fe}}{m_i}\nabla (\frac{n}{n_0})^{2/3}+\frac{\hbar^2}{
2m_e m_i}\nabla \frac{1}{\sqrt{n}}\Delta \sqrt{n} \ ,
\end{eqnarray}
where $u(\vec{r},t)$ is the macroscopic velocity of the plasma
\begin{eqnarray}
\label{mvel}
\vec{u}=\frac{1}{n}\int \frac{2d^3p}{(2\pi \hbar )^3}\vec{v}f(\vec{r},\vec{p},t)\ .
\end{eqnarray}
Obviously from Eqs.(\ref{con}) and (\ref{mot}) after linearization follows the
Bogolyubov's type of dispersion equation (\ref{bog}).

The more general set of fluid equations for $\alpha $
kind of particles we can obtain from the equation (\ref{tob}) taking into account that any elastic scattering
should fulfill the general conservation laws of particle number, momentum and
energy
\begin{eqnarray}
\label{gcon}
\frac{\partial n_\alpha }{\partial t}+\nabla \cdot(n_\alpha \vec{u}_\alpha )=0\ ,
\end{eqnarray}
\begin{eqnarray}
\label{gmot}
\frac{\partial <\vec{p}_\alpha >}{\partial t}+(\vec{u}_\alpha \cdot
\nabla ) <\vec{p}_\alpha >=e_\alpha \Bigl( \vec{E}+\frac{1}{c}\vec{u}_\alpha\times\vec{B}\Bigr) -
\frac{1}{n_\alpha }\nabla P_\alpha +
\frac{\hbar^2}{2m_\alpha }\nabla \frac{1}{\sqrt{n_\alpha }}\Delta
\sqrt{n_\alpha }  \nonumber \\
+\frac{1}{n_\alpha }\int \frac{2dp}{(2\pi \hbar )^3}\ p_\alpha\ C(f_\alpha)\ ,
\end{eqnarray}
where $P_\alpha=\frac{1}{3m_\alpha}\int\frac{d^3p}{(2\pi\hbar)^3}\frac{p_\alpha^2}{exp\{\frac{\varepsilon_\alpha-\mu_\alpha}{
T_\alpha}\}+1}\ .$

Moreover, the equation of the internal energy of each particle species can be derived from the
kinetic equation (\ref{tob}) by multiplying it by $\frac{p_\alpha^2}{
2m_\alpha },$ integrating over the momentum and employing the equations of
continuity (\ref{gcon}) and motion (\ref{gmot}). In the non-relativistic limit the result is
\begin{eqnarray}
\label{state}
\frac{d}{dt}\ln \frac{<\varepsilon_\alpha >}{n_\alpha^{2/3}}=\frac{1
}{n_\alpha <\varepsilon_\alpha >}\int \frac{2d^3p}{(2\pi
\hbar )^3}\ \varepsilon_\alpha \ C(f_\alpha )\ ,
\end{eqnarray}
where $\varepsilon_\alpha =\frac{(p_\alpha-<p_\alpha >)^2}{2m_\alpha }$ and $
<\varepsilon_\alpha >=\frac{1}{n_\alpha }\int \frac{2d^3p}{
(2\pi \hbar )^3}\ \varepsilon_\alpha f_\alpha $ is the internal
energy of particles, and the collision terms can be positive or negative.

\section{Derivation of the Korteweg-de Vries Equation }

We now consider the propagation of weakly nonlinear ion-sound waves. To this end, we generalize the Bogolyubov's type of dispersion relation (\ref{bog}), where we assumed the quasi-neutrality (\ref{qneu}),  taking into account the small separation of charges. For the linear ion-sound waves ($kv_{Fe}\gg\omega\gg kv_{Fi}$) we have
\begin{eqnarray}
\label{lind}
\omega =kv_s+\beta k^3-\alpha k^3 \ ,
\end{eqnarray}
where $v_s=\sqrt{m_e/3m_i}v_{Fe}\ $ is the ion sound velocity, $\ \beta=\frac{3}{8}\lambda_B^2v_s,\ $ and $\ \alpha=v_s\lambda_{TF}^2/3\ $ is due to the separation of charges.

Equation (\ref{lind}) describes sufficiently long waves in plasmas, where $\omega/k$ at $k\rightarrow 0$ has a finite limit (weakly dispersive waves). The nonlinear waves for the weakly dispersive plasmas are governed by the Korteweg-de Vries equation, which we shall derive in the following.

As is well known the nonlinearity, generally speaking, causes the distortion of wave profile, which steepens until the dissipative effects come into play. The later causes the broadening of the profile and finally balances the nonlinear steepening, so that the nonlinear stationary waves can be formed in the dispersive plasma. To show this, we follow the standard method \cite{whi}-\cite{kar}, and from Eq.(\ref{del}) obtain the expression for the density of electrons
\begin{eqnarray}
\label{elde}
n_e=n_i+\frac{3}{2}n_0\lambda_{TF}^2\Delta\Bigl(\frac{n_i}{n_0}\Bigr)^{2/3} \ ,
\end{eqnarray}
where $n_0=n_{0e}=n_{0i}$ is the unperturbed density. Note that the second term in Eq.(\ref{elde}) is small since $\lambda_{TF}^2\ll R^2$, where R is the characteristic scale length of the perturbation.

In one dimensional case equations (\ref{con}) and (\ref{mot}) now read
\begin{eqnarray}
\label{odcon}
\frac{\partial n_i}{\partial t}+\frac{\partial}{\partial x}n_iu=0
\end{eqnarray}
\begin{eqnarray}
\label{odmot}
\frac{\partial u}{\partial t}+u\frac{\partial}{\partial x}u=-\frac{\varepsilon_{Fe}}{m_i}
\frac{\partial}{\partial x}\Bigl(\frac{n_e}{n_0}\Bigr)^{2/3}+\frac{\hbar^2}{2m_em_i}\frac{\partial}{\partial x}
\frac{1}{\sqrt{n_e}}\Delta\sqrt{n_e}\ .
\end{eqnarray}
Substitution of Eq.(\ref{elde}) into Eq.(\ref{odmot}) yields the equation of motion for ions
\begin{eqnarray}
\label{emio}
\frac{\partial u}{\partial t}+u\frac{\partial}{\partial x}u=-\frac{3}{2}v_s^2\frac{\partial}{\partial x}\Bigl(
1+\lambda_{TF}^2\frac{\partial^2}{\partial x^2}\Bigr)\Bigl(\frac{n_i}{n_0}\Bigr)^{2/3}+
\frac{\hbar^2}{2m_em_i}\frac{\partial}{\partial x}
\frac{1}{\sqrt{n_i}}\frac{\partial^2}{\partial x^2}\sqrt{n_i}\ .
\end{eqnarray}
Here the last two terms describe the dispersive effects.

Obviously from Eqs.(\ref{odcon}) and (\ref{emio}), which formally resemble the Boussinesq equations, after the linearization follows the dispersion equation (\ref{lind}). If we neglect the dispersive terms in Eq.(\ref{emio}), then Eqs.(\ref{odcon}) and (\ref{emio}) reduce to the Euler equations.

We now consider the nonlinear one dimensional traveling waves in a single  direction, so-called simple waves. Since the velocity u and the density $n_i$ in such a wave are functions of the same variable $x-ut$, we suppose that for a wave with any amplitude, all quantities describing a simple wave are represented in the form of functions of one of these quantities, i.e. it can be assumed that $u=u(n)$ or $n=n(u)$. From Eqs.(\ref{emio}) (without the dispersive terms) and (\ref{odcon}) then follows for a wave propagating in the positive x direction
\begin{eqnarray}
\label{wpod}
\frac{du}{dn_i}=\frac{v_s}{n_0^{1/3}}\frac{1}{n_i^{2/3}} \hspace{1cm} or \hspace{1cm}
u=3v_s\left[\Bigl(\frac{n_i}{n_0}\Bigr)^{1/3}-1\right]
\end{eqnarray}
and
\begin{eqnarray}
\label{wpand}
x=t\left[v_s+\frac{4}{3}u\right]+F(u) \ ,
\end{eqnarray}
where $F(u)$ is an arbitrary function of the velocity.

Expressions (\ref{wpod}) and (\ref{wpand}) are general solutions of arbitrary traveling waves and determine the velocity as an implicit function of x and t.

We can also rewrite Eq.(\ref{wpand}) in the form
\begin{eqnarray}
\label{resev}
u=Q(x-v(u)t)\ , \hspace{1cm} v(u)=v_s+\frac{4}{3}u\ ,
\end{eqnarray}
where Q(x) is an arbitrary function defining the wave shape at t=0
\begin{eqnarray*}
Q(x)=u(x,0) \ .
\end{eqnarray*}

Since the velocity u is the function of density, therefore it is different for different points at the wave profile, i.e., the profile changes in the course of time. One can see from Eq.(\ref{wpod}) that $du/dn>0$, i.e., the velocity of propagation of a given point at the wave profile increases with the density. This condition $du/dn>0$ is of the shock waves \cite{whi}-\cite{kar}. At the point, where the shock is formed, the dispersive terms that we have neglected in Eq.(\ref{emio}) become important because the nonlinear steepening can be compensated by the dispersive spread. For some conditions the formation of solitary waves takes place.

Expressing the ion density through the velocity from Eq.(\ref{wpod}) and substituting it into Eq.(\ref{emio}), we obtain KdV equation
\begin{eqnarray}
\label{kdv}
\frac{\partial u}{\partial t}+\Bigl(v_s+\frac{4}{3}u\Bigr)\frac{\partial u}{\partial x}=-(A-B)\frac{\partial^3 u}{\partial x^3}\ ,
\end{eqnarray}
where $A=v_s\lambda_{TF}^2$ and $B=\hbar^2/4m_em_iv_s$.

We emphasize here that Eq.(\ref{kdv}) has two physically distinct solutions. Namely, if the right-hand side of this equation is negative, then the solution of Eq.(\ref{kdv}) corresponds to the compressional soliton. Whereas, in the opposite case, when the quantum (Madelung) term exceeds the term due to the charge separation, the solitary wave is a rarefaction wave.

Interesting case is when $A\approx B$. In this case there is no more balance between the terms of the charge separation and the quantum, and the nonstationary case of shock formation takes place. The condition $A\approx B$ is satisfied for the density $n\sim 10^{25}cm^{-3}$.

\section{Summary}

Starting with the Schr\"{o}dinger equation for  a single  particle, we have derived the Wigner-Moyal equation and pointed out the limitations of the usability of this equation. Namely, we have revealed that the Wigner quantum kinetic equation is not appropriate for the construction of fluid equations. It is elucidated that the Wigner distribution function has no physical meaning. We have also demonstrated that the Wigner-Moyal type of equation  can be derived for any system, classical or quantum. We have obtained a new type of quantum kinetic equation motivated by its necessity for the correct description of the collective behavior of the Fermi quantum plasma. This novel quantum kinetic equation allows to obtain a set of quantum hydrodynamic equations. In our formulation of kinetic, as well as hydrodynamic equations the quantum term, which contains all the information on the quantum effects, is incorporated. We have study the propagation of small longitudinal perturbations in an electron-ion collisionless plasmas, deriving a quantum dispersion equation. We have discussed the quantum correction to the zero sound. For the special interest case we have derived a quantum kinetic equation in the one-fluid approximation and have shown that, when a charge and the Thomas-Fermi screening length are not present in the equation, the solution of linear kinetic equation leads to the Bogolyubov's type of dispersion equation, which is valid in the Bose fluid. It is clear that in the dense strongly coupled plasmas with attractive interaction the formation of the bound state is possible, so that a neutral Bose atoms can be created.
Furthermore, we have derived KdV equation in order to discuss the nonlinear effects in a quantum plasma, and found that due to the quantum term the compressional solitons may become rarefaction waves.
These investigations may play an essential role for the description of complex phenomena that appear
in dense astrophysical objects, as well as in the next generation intense laser-solid density plasma experiments.

\end{document}